# An optical fan for light beams for high-precision optical measurements and optical switching

Zhi-Yuan Zhou[1,2,#1], Yan Li[1,2,#1], Dong-Sheng Ding[1,2], Yun-Kun Jiang[3], Wei Zhang[1,2], Shuai Shi[1,2], Bao-Sen Shi[1,2,*] and Guang-Can Guo[1,2]

[1]*Key Laboratory of Quantum Information, University of Science and Technology of China, Hefei, Anhui 230026, China*

[2]*Synergetic Innovation Center of Quantum Information & Quantum Physics, University of Science and Technology of China, Hefei, Anhui 230026, China*

[3]*College of Physics and Information Engineering, Fuzhou University, Fuzhou 350002, China*

[*]*Corresponding author: drshi@ustc.edu.cn*

The polarization and orbital angular momentum properties of light are of great importance in optical science and technology in the fields of high precision optical measurements and high capacity and high speed optical communications. Here we show, a totally new method, based on a combination of these two properties and using the thermal dispersion and electro-optical effect of birefringent crystals, the construction of a simple and robust scheme to rotate a light beam like a fan. Using a computer-based digital image processing technique, we determine the temperature and the thermal dispersion difference of the crystal with high resolution. We also use the rotation phenomenon to realize thermo-optic and electro-optic switches. The basic operating principles for measurement and switching processes are presented in detail. The methods developed here will have wide practical applicability in various fields, including remote sensing, materials science and optical communication networks.

The freedoms of the properties of light, such as its frequency, polarization and spatial modes, play vital roles in optical science and technology. The ability to control these properties of light determines one's ability to progress in the field of optics. The spatial freedom of the orbital angular momentum (OAM) of light has drawn a great deal of attention since it was first introduced by Allen et al. [1]. The spatial structure of light with OAM makes it preferable for optical trapping and optical tweezers applications [2, 3]. Because there is no dimensional limitation for OAM, a light beam carrying OAM can enhance the transmission capacity in both free space and fiber optic communications [4, 5]. Light with OAM is also widely studied in quantum optics to demonstrate the basic principles of quantum mechanics [6–10], to create a high-dimensional entanglement [7, 10], to enhance the security and the information capacity

---

[#1] **These two authors have contributed equally to this article.**

of quantum key distribution in free space [11–13], and to improve both the resolution in quantum imaging [14] and the measurement precision in quantum metrology [15]. By applying light beams carrying OAM to third- and second-order nonlinear processes such as four-wave mixing in atomic vapors [16–19] and second harmonic generation in nonlinear crystals [20–23], some new phenomena and physics are revealed in comparison with the effects when using Gaussian light.

Optical measurements [24] offer the benefits of high precision and non-contact processes, and the resolution of optical measurements is usually orders of magnitude better than that of other methods. The non-contact property makes optical measurements very useful in certain extreme conditions, such as high temperature or high pressure scenarios. Optical measurements are especially important in remote sensing applications, including measurement of the physical parameters of biological tissues [25], night vision and satellite remote sensing. Additionally, the optical switch is a basic element in optical communication networks for the multiplexing and demultiplexing of optical signals. The most commonly used switching methods include Mach–Zehnder interferometer-based switches and wavelength switching based on gratings. Robustness, high switching speeds and high numbers of switchable paths are very important for optical switches.

In this article, a totally new method for high precision optical measurements and optical switching is demonstrated, using a combination of a light beam carrying OAM and a birefringent crystal, we have realized a 'fan-like' optical rotator for the OAM-carrying light beam by tuning the crystal temperature or electro-optical effect. By applying a computer-based digital image processing technique, we can determine both the temperature and the thermal dispersion difference of the crystal with high resolution. The simplicity and robustness of our experimental setup mean that it is very promising for practical temperature sensing applications and for measurement of the thermal dispersion difference of birefringent materials. Additionally, the rotation of the beam with OAM can also be used for optical switching; multiple paths can be switched simultaneously, depending on the value of the OAM carried by the light beam in our switching scheme. The experimental setup and the basic operating principles of our experiments are described in detail in the following.

The experimental scheme is depicted in Figure 1. We use quantum mechanical language to describe the transformation of the light. The beam from the laser (Ti:sapphire, MBR110, Coherent) passes through a Sagnac interferometer to generate a hyper-superposition state of polarization and OAM in the form of [10, 13]

$$|\Phi\rangle_1 = \frac{1}{\sqrt{2}}(|H\rangle|l\rangle + e^{i\alpha}|V\rangle|-l\rangle) \qquad (1)$$

Here, $|H\rangle$ and $|V\rangle$ denote the horizontal and vertical polarizations of the light, respectively. $|l\rangle$ represents the OAM degree of freedom of the light, which is created using a vortex phase plate (VPP). $\alpha$ denotes the phase, which depends on the angular positions of a half wave plate (HWP) and a quarter wave plate (QWP) located at the input port of the interferometer. The light from the output port of the Sagnac interferometer is focused into a potassium titanyl phosphate crystal (KTP); the crystal has been x-cut for light propagation along the x-axis from the crystal's principal axes. The temperature of the crystal is controlled

using a homemade semiconductor Peltier cooler with temperature stability of ±2 mK. After leaving the crystal, the light polarization is rotated by a HWP, projected into the horizontal polarization by a polarizing beam splitter (PBS) and imaged using a charge coupled device (CCD) camera. The image acquired by the CCD is processed using a digital image processing technique for further analysis.

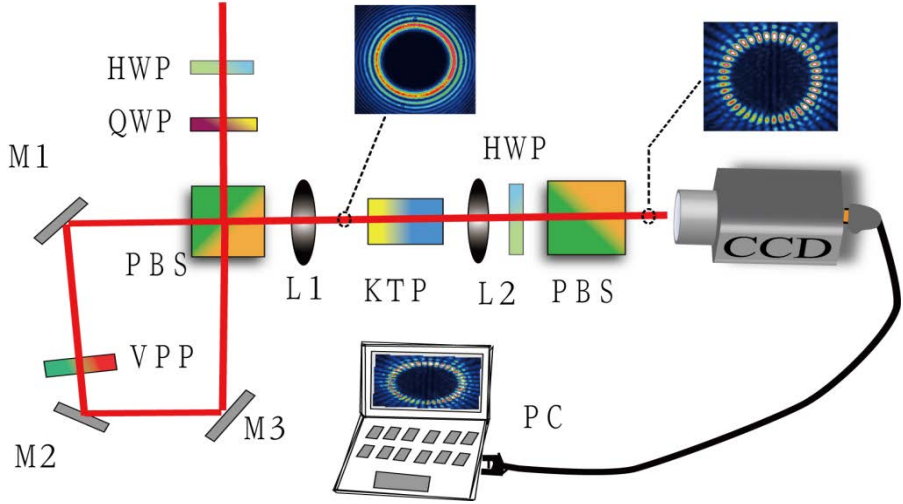

Figure 1. Experimental setup. HWP: half wave plate; QWP: quarter wave plate; L1, L2: lenses; PBS: polarizing beam splitter; M1–M3: mirrors; VPP: vortex phase plate; PC: personal computer; CCD: charge coupled device camera; KTP: potassium titanyl phosphate crystal.

The operational principle of our thermal rotator is as follows: after a light beam passes through the KTP crystal, the H and V polarizations of the light in equation (1) experience different optical paths, and thus the state of the beam after passing through the KTP crystal is

$$|\Phi\rangle_2 = \frac{1}{\sqrt{2}}(|H\rangle|l\rangle + e^{i(\alpha+\beta)}|V\rangle|-l\rangle), \qquad (2)$$

where $\beta = 2\pi(n_z(\lambda,T) - n_y(\lambda,T))L/\lambda$ is the optical path difference caused by the birefringence of the crystal. $n_z$ and $n_y$ are the refractive indexes of the KTP crystal, $\lambda$ is the laser wavelength, $T$ is the crystal temperature, and $L$ is the length of the KTP crystal. After the light passes through the HWP (which has optical axes at 22.5 degrees relative to the horizontal) and the PBS, its state becomes

$$|\Phi\rangle_3 = \frac{1}{\sqrt{2}}|H\rangle \otimes (|l\rangle + e^{i(\alpha+\beta)}|-l\rangle). \qquad (3)$$

State $|\Phi\rangle_3$ is the superposition of two different OAMs, and generates an interference pattern containing $2l$ petals arranged symmetrically in a ring. The pattern will rotate with changes in $\beta$. When $\beta$ changes by $2\pi$, the pattern rotates by an angle of $\pi/l$. A continuous increase in temperature will make the pattern rotate continuously in the same direction, similar to a mechanical fan. When the wavelength remains unchanged and the crystal temperature is tuned, then the phase variance $\Delta\beta = 2\pi L\left(\partial n_z(\lambda,T)/\partial T - \partial n_y(\lambda,T)/\partial T\right)\Delta T/\lambda$ depends on the temperature variance. Therefore, the relative rotation angle of the pattern is related to the relative change in temperature by the following formula:

$$\Delta\theta = \frac{\pi L}{l\lambda}\left(\frac{\partial n_z(\lambda,T)}{\partial T} - \frac{\partial n_y(\lambda,T)}{\partial T}\right)\Delta T. \qquad (4)$$

Equation (4) shows that when we can measure the pattern rotation angle with high resolution, we can then determine the temperature or the thermal dispersion difference with high precision. The total rotation angle depends on the total temperature change, and the rotation speed depends on the rate of temperature change.

Figure 2 shows the rotation angle versus the crystal temperature with OAM values of 2 and 20. Figure 2a and b show the interference patterns for $l = 2$ and 20, respectively, and the numbers of petals can be clearly distinguished. Figure 2c and d show the relationships between the rotation angle and the relative change in temperature for $l = 2$ and 20, respectively. There are two data sets in both Figure 2c and d, corresponding to the two different digital image processing methods that were used [26]. The basic ideas for determination of the rotation angle consist of two elementary operations on two images (where one is the reference image, and the other is the target image): the first operation is image rotation, and the second operation is image correlation. By scanning the angle of rotation of the reference image with small steps and calculating the value of the correlation with the target image, we can obtain a correlation curve as a function of the rotation angle, where the value when the correlation value reaches a maximum is the angle of rotation between the reference and target images. Figure 2f shows a typical correlation curve for the two images used in our numerical calculations for $l = 20$; we found that the correlation curve varies periodically with the rotation angle, because of the periodic distribution of the interference pattern in the azimuthal direction. The data sets shown in Figure 2c and d are obtained by tuning the crystal temperature from $T_0$ to $T_n$ and recording the corresponding image for processing. In the calculation of data set $G_n\_G_0$, the reference image is the image at the initial temperature $T_0$ and the target image is the image at temperature $T_n$, while for the calculation of data set $G_n\_G_{n-1}$, the reference image is the image at the previous temperature $T_{n-1}$ and the target image is the image at temperature $T_n$. The slopes of the fitted lines in Figure 2c and d determine the sensitivity of the angle relative to the crystal temperature. The slopes in Figure 2c and d are 7.795 degrees/K (7.772 degrees/K) and 0.8151 degrees/K (0.8210 degrees/K) for data set $G_n\_G_0$ ($G_n\_G_{n-1}$), respectively. The small differences in slope between the two calculation methods stem from the divergence in determination of the center of the reference image for rotation. The numerically calculated angular resolution of our image processing program (written in C language) is 0.01 degrees, which corresponds to a temperature resolution of $1.28\times10^{-3}$ K ($1.23\times10^{-2}$ K) for $l = 2$ (20). The numerically calculated resolution for the rotation angle can be further improved by optimization of the numerical algorithm used to determine the center of the reference image for rotation. When we know the rotation angle and the temperature variance, we can then determine the thermal dispersion difference of the crystal. The value of $\frac{\partial n_z(\lambda,T)}{\partial T}-\frac{\partial n_y(\lambda,T)}{\partial T}$ measured by our method is $7.252\times10^{-6}$, while the theoretical value calculated according to Ref. 26 is $1.0701\times10^{-5}$; these two values are rather different, but we believe that our measurement is much more reliable, because the experimental setup is very stable and robust. We also characterized the temperature tuning property by filtering out a single petal of the interference pattern using a small aperture, and measured the leakage power using a power meter. The result for $l=2$ is shown in Figure 2e, and indicates than the leakage power depends periodically on the crystal temperature.

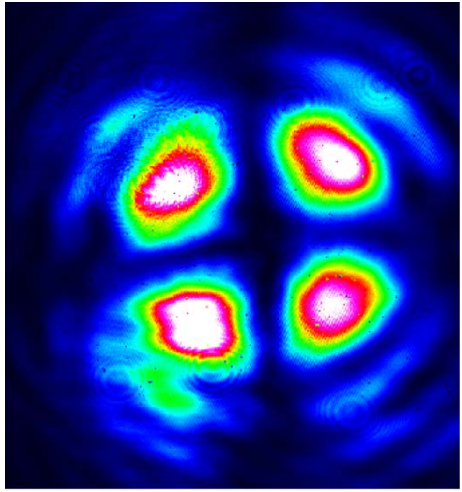
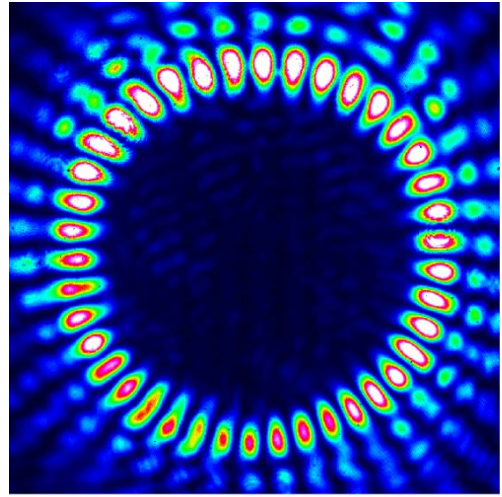

a                                    b

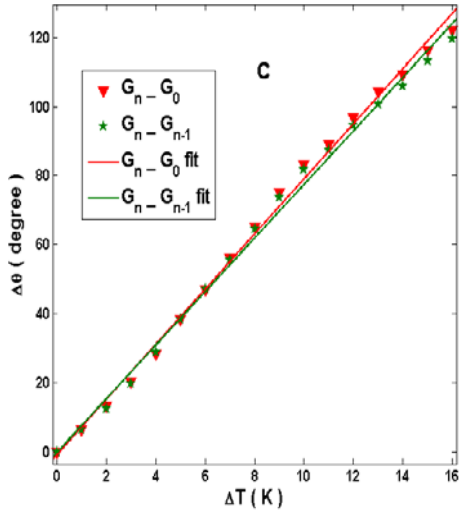
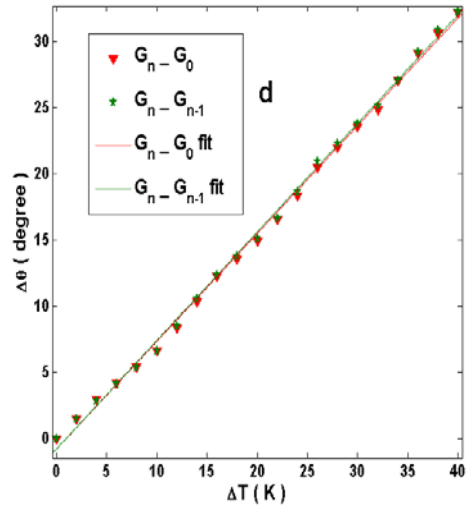

c                                    d

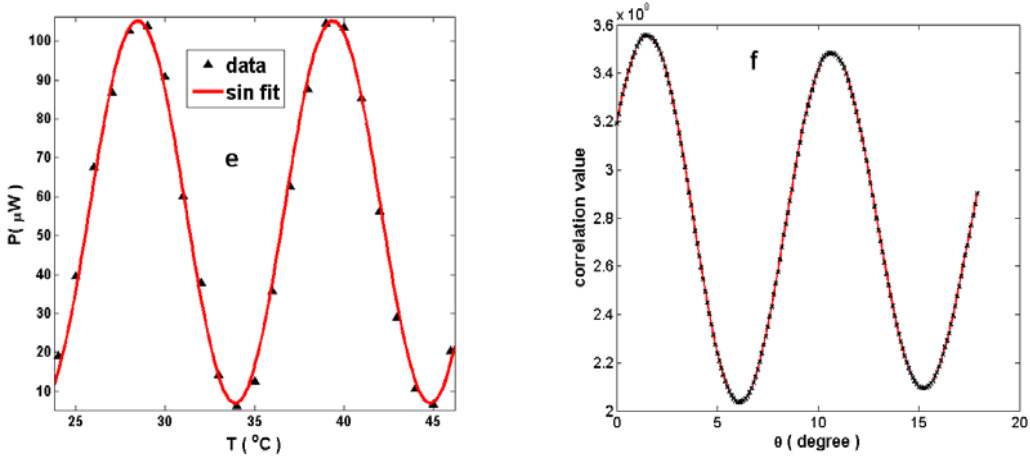

Figure 2. Experimental data. a and b show the interference patterns for $l = 2$ and 20, respectively; c and d show the relative rotation angle as a function of the relative temperature change of the crystal for $l = 2$ and 20, respectively; e shows the leakage power from the aperture as a function of the crystal temperature for $l = 2$; f shows a typical correlation curve from the numerical calculation for $l = 20$.

In addition to performing precision measurements using this experimental setup, the rotation property can also be used to form a thermo-optic switch. The operating principle of our switch is as follows. The interference pattern has an angular spacing of $\pi/l$, which is dependent on the value of the OAM of the light beam. By dividing the angular spacing equally into N spatial points with angular spacing of $\pi/lN$, we can then scan the petal through these spatial points one by one. By encoding information via the input port of a Gaussian beam, the $2l$ petals will have the same code information, and we can therefore switch between the $2l$ users in a group simultaneously. We defined the full wave temperature $T_F$ for our switch, corresponding to an angular rotation of $\pi/l$. From equation (4), we obtained $T_F = \lambda / L(\frac{\partial n_z(\lambda,T)}{\partial T} - \frac{\partial n_y(\lambda,T)}{\partial T})$. In our experiments, $T_F$ is 10.96K. The switching temperature between two adjacent spatial points is $T_F / N$.

The speed of thermo-optic switching is usually slow, but we can realize fast switching using the electro-optical effect of a birefringent crystal. We used an electro-optic modulator (EOM, model 4002, New Focus) to realize fast rotation of the interference pattern. In this experiment, the KTP crystal is replaced with the EOM. The EOM is controlled by a high voltage amplifier connected to a signal generator. We use a small aperture to filter out only part of a single petal, and detect the leakage light beam using a fast photodiode; the photodiode signal is then input to an oscilloscope. By applying a triangular signal to the EOM, the pattern can be rotated back and forth. The voltage scanning signal and the signal from the detector are shown in Figure 3. Here, the EOM scanning frequency is 100 Hz, while the output range of our high voltage amplifier is smaller than the full-wave voltage of the EOM, so the detector signal only shows part of a whole period. The scanning frequency can reach MHz by application of a high-speed high voltage source, which ensures the realization of fast switching using our switching method.

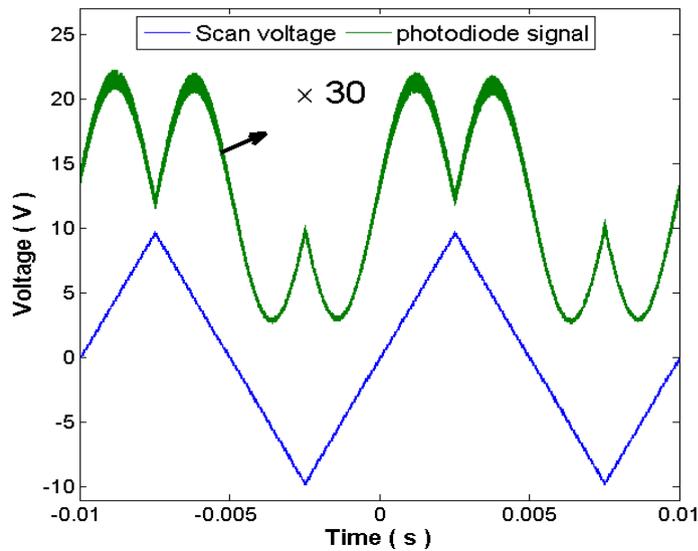

Figure 3. Voltage scanning signal of the EOM and the intensity signal from the photodiode detector, recorded using the oscilloscope. The scanning amplitude is attenuated by 20 times before connection to the oscilloscope, while the detector signal shown in the figure is 30 times the magnitude of the real signal.

In summary, by combination of the OAM and polarization properties of a light beam, and using birefringent crystals, we have realized an optical fan-like phenomenon using a light beam carrying OAM. Using a digital image processing technique, we have determined the temperature and the thermal dispersion difference of the birefringent crystal with high resolution, which is very promising for practical materials science and remote temperature sensing applications. The rotation property can also be used to realize both thermo-optic and electro-optic spatial switching, and is potentially of great importance for applications in optical communications.

**Acknowledgements**

This work was supported by the National Fundamental Research Program of China (Grant No. 2011CBA00200), the National Natural Science Foundation of China (Grant Nos. 11174271, 61275115, 10874171), and the Innovation Fund from the Chinese Academy of Science.